\documentclass[epjCONF,columns]{svjour} 
\usepackage{graphics}
\usepackage[varg]{txfonts} 
\usepackage[latin1]{inputenc}
\session-title{Hadron Collider Physics symposium 2011}
\begin{document}
\title{Measurement of the ttbar production cross section in the fully hadronic decay channel in pp collisions at 7 TeV
}
\author{A. Tropiano\inst{1}\fnmsep\thanks{on behalf of the CMS Collaboration}}
\institute{Universit\`a degli Studi di Firenze,\\ \email{antonio.tropiano@cern.ch}}
\abstract{
The first measurement of the top quark pair production cross section in the fully hadronic channel at a center-of-mass energy of $\sqrt s = 7$ TeV is presented. The measurement has been performed using an integrated luminosity of 1.09 fb$^{-1}$, collected with the CMS detector. The reconstruction of the $t\bar t$ candidates is performed after a cut-based event selection using a kinematic fit. A data-driven technique is used to estimate the dominant background from QCD multijet production. The cross section is determined from a fit to the top quark mass. The cross section measurement yields $\sigma(t\bar t)=136 \pm 20 (stat.) \pm 40 (sys.) \pm 8 (lumi.)$ pb. This result is consistent with an independent measurement in the same channel, with the measurements in other decay channels and with the Standard Model predictions.
} 
\maketitle
\section{Introduction}
\label{intro}
Measurements of the top quark pair production cross section in proton-proton collisions at the LHC provide important tests of our understanding of the top quark production mechanism and can also be used in new physics searches. 
Recent CMS \cite{cms} measurements, making use of the full 2011 dataset corresponding to an integrated luminosity greater than 1 fb$^{-1}$, have been performed in the dilepton  channel, including $\tau$ leptons, and in the lepton+jets channel \cite{dilepton,tau,semilepton}.
 
Here the first cross section measurement in the fully hadronic decay channel with the CMS detector is presented \cite{hadronic}. The fully hadronic decay channel has the largest signal yield compared to the dilepton and lepton+jets channels. However, with only jets in the final state, this channel is dominated by the nearly identical QCD multijet background. This measurement demonstrates the good understanding of the  CMS detector and is, therefore, an important milestone. This measurement was combined with the more recent measurements in the other channels, providing the most precise CMS measurement obtained to date.
%
\section{Event Selection}
To select fully hadronically decaying top quark pairs the candidate events are required to have at least four jets with $p_T > 60$ GeV/c, a fifth jet with $p_T > 50$ GeV/c, and a sixth jet with $p_T >$ 40 GeV/c. All jets are required to be within an absolute pseudorapidity $|\eta|$ of 2.4, in the tracker acceptance. Additional jets are considered only if they have a transverse momentum of more than 30 GeV/c.

Tagging jets originating from bottom quarks ($b$-tagging) is achieved with a robust $b$-tagging algorithm based on the reconstruction of secondary vertices \cite{btag}. This algorithm yields an efficiency of 38$\pm$4\% while having a mis-tag rate of 0.12$\pm$0.02\%. At least two b-tagged jets are required in each event.

\subsection{Kinematical Fit}
For the final selection of candidate top quark events a kinematic (least-squares) fit is applied. It exploits the characteristic topology of top quark events: two $W$ bosons with a mass of 80.4 GeV/c$^2$ that can be reconstructed from the untagged jets and two top quarks that can be reconstructed from the $W$ bosons and the $b$-tagged jets. The masses of the two top quarks are assumed to be equal, but not fixed, in order to avoid any bias.

To find the right combination of jets, the fit procedure is repeated for every distinguishable jet permutation. This is done using all (six or more) jets that passed the selection criteria. All $b$-tagged jets are taken as bottom quark candidates, the remaining jets are taken as light quark candidates. If the fit converges for more than one of the possible jet permutations, the one with the minimum $\chi^2$ is chosen. After the kinematic fit, all events with a fit probability $P(\chi^2) >$ 0.01 are accepted. In the kinematic fit gaussian resolutions are used for the jets. 

The number of events in data passing each selection step and the expected signal fraction using a Standard Model $t\bar t$  production cross section of 163 pb \cite{kidonakis} are given in Table \ref{tab:selection}.

\begin{table}[ht!]
\begin{center}
\caption{Number of events and the expected signal fraction in the data sample after each selection step. The expected signal fraction is taken from the simulation, assuming a cross section of 163 pb.}
\label{tab:selection}       
\begin{tabular}{lll}
\hline\noalign{\smallskip}
Selection Step & Events & Signal Fraction  \\
\noalign{\smallskip}\hline\noalign{\smallskip}
At least 6 jets  & 248109 & 2\%\\
At least 2 $b$-tags & 6905 & 17\% \\
Kinematical Fit & 1620 & 32\% \\
\noalign{\smallskip}\hline
\end{tabular}
\end{center}
\end{table}

\section{Signal Extraction}
\begin{figure}[ht!]
\begin{center}
\resizebox{0.75\columnwidth}{!}{\includegraphics{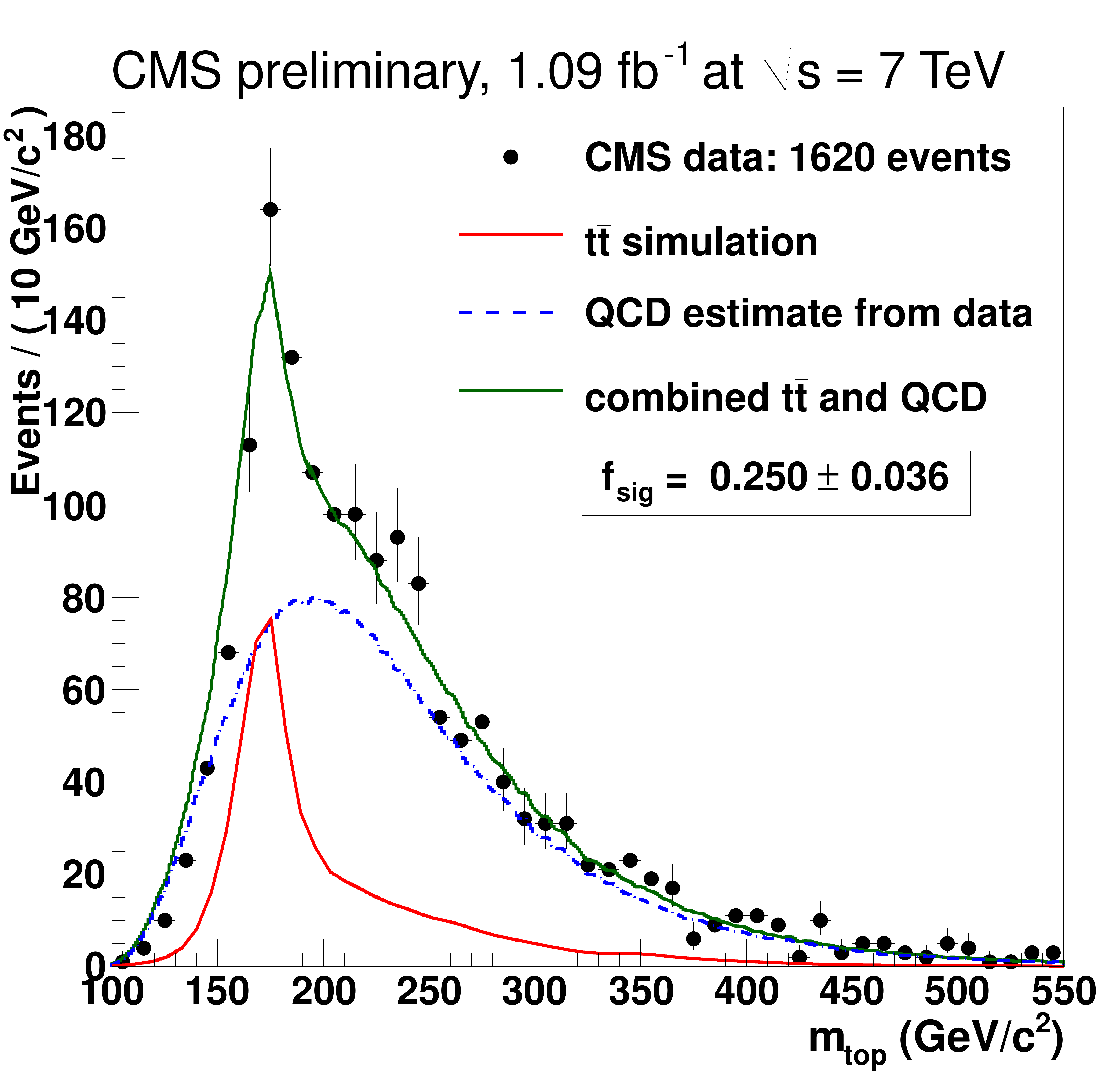}}
\caption{Result of the fit of the reconstructed top quark mass for the $t\bar t$ simulation (solid line) and the multi-jet QCD estimated from data (dashed line). The uncertainty stated on the signal fraction f$_{sig}$ is only statistical.}
\label{fig:topmass}       
\end{center}
\end{figure}

The number of signal events after the final selection is determined via an unbinned maximum likelihood fit to the reconstructed top quark mass distribution, obtained from the kinematic fit. The shapes used in the fit for the signal and background distributions are derived from simulation and a data-driven estimate, respectively. The resulting distribution is shown in Figure \ref{fig:topmass}.

\section{QCD Background Estimation}
The background from QCD multijet events is estimated from events with six or more jets of which exactly zero are $b$-tagged. In this region of the phase space the signal contribution is below 1\%. As the kinematics of $b$-tagged jets and non-$b$-tagged jets is different, the events that do not contain any b-tag are weighted such that they reproduce the kinematics of the $b$-tagged jets. This is done by deriving a ratio:
\begin{equation}
R(p_T, |\eta|) = \frac{N(p_T, |\eta|, b)}{N(p_T, |\eta|, not ~b)}                      
\end{equation}
where $N(p_T, |\eta|, b)$ is the number of $b$-tagged jets in events with at least one jet coming from $b$ quarks and\\ $N(p_T, |\eta|, not ~b)$ is the number of untagged jets from all events in data. The ratio $R$ gives the probability of  $b$-tagging a jet as a function of its transverse momentum $p_T$ and absolute pseudorapidity $|\eta|$. The kinematic fit is performed on all events with zero $b$-tagged jets, assuming every jet as bottom quark candidate. For all permutations with $P(\chi^2) \geq 0.01$ an event weight
\begin{equation}
w = R(p_T^b , |\eta|^b) \times R( p_T^{\bar b} , |\eta|^{\bar b})                              
\end{equation}
is calculated depending on the $p_T$ and $|\eta|$ of the two jets assigned to the bottom quarks in the kinematic fit hypothesis. The weighted events are used for the estimation of the QCD multijet background that passes the event selection criteria with two $b$-tagged jets.

\section{Results}
The inclusive cross section for top quark pair production in the fully hadronic decay channel has been derived from
\begin{equation}
\sigma_{t\bar t} = \frac{f_{sig} \cdot N}{\epsilon \cdot L_{int}}                                                                                   
\end{equation}
The fraction of signal events, f$_{sig}$, is estimated by the fit of the top quark mass distribution and is found to be 25.0\%. The number of events $N$ seen in the data after the full selection is 1620 and the efficiency for selecting $t\bar t$ events $\epsilon$, which is determined from simulation, is 0.22\%. The latter includes the correction factors for the b-tagging efficiency, mis-tag rate and trigger efficiency found in data. The integrated luminosity of the data sample is L$_{int}$ = 1.09 fb$^{-1}$. With these numbers the $t \bar t$ production cross section $\sigma_{t \bar t}= 136 \pm 20 (stat.) \pm 40 (sys.) \pm 8 (lumi.)$ pb
is obtained. The total relative uncertainty corresponds to $\pm$33\%.
 
To determine the systematic uncertainties on the cross section measurement, samples of simulated events with modified parameters have been used. For each variation the change in the efficiencies and the fitted signal fraction, and hence the cross section, has been studied using pseudo-experiments. An overview of the main sources of systematic uncertainties is given in Table \ref{tab:systematics}.
\begin{table}[ht!]
\begin{center}
\caption{Summary of the main sources of systematic uncertainties with their values.}
\label{tab:systematics}       
\begin{tabular}{lll}
\hline\noalign{\smallskip}
Systematic Error & Relative Uncertainty\\
\noalign{\smallskip}\hline\noalign{\smallskip}
$b$-tagging & 15.7\%\\
Jet Energy Scale & 13.5\%\\
Background & 12.2\%\\
Q$^2$ scale & 8.7\%\\
\noalign{\smallskip}\hline
\end{tabular}
\end{center}
\end{table}
\section{Cross Check Analysis}
A second measurement has been performed as a cross check. For this analysis the kinematic properties of signal and background events are characterized to develop a kinematic selection based on a neural network. Moreover, the QCD multijet background is estimated from data using the information from the kinematic variables of dijet pairs that passed loose $b$-tagging criteria and extrapolating to the signal region with tighter $b$-tagging criteria and higher jet multiplicities. 

\begin{figure}[ht!]
\begin{center}
\resizebox{0.75\columnwidth}{!}{\includegraphics{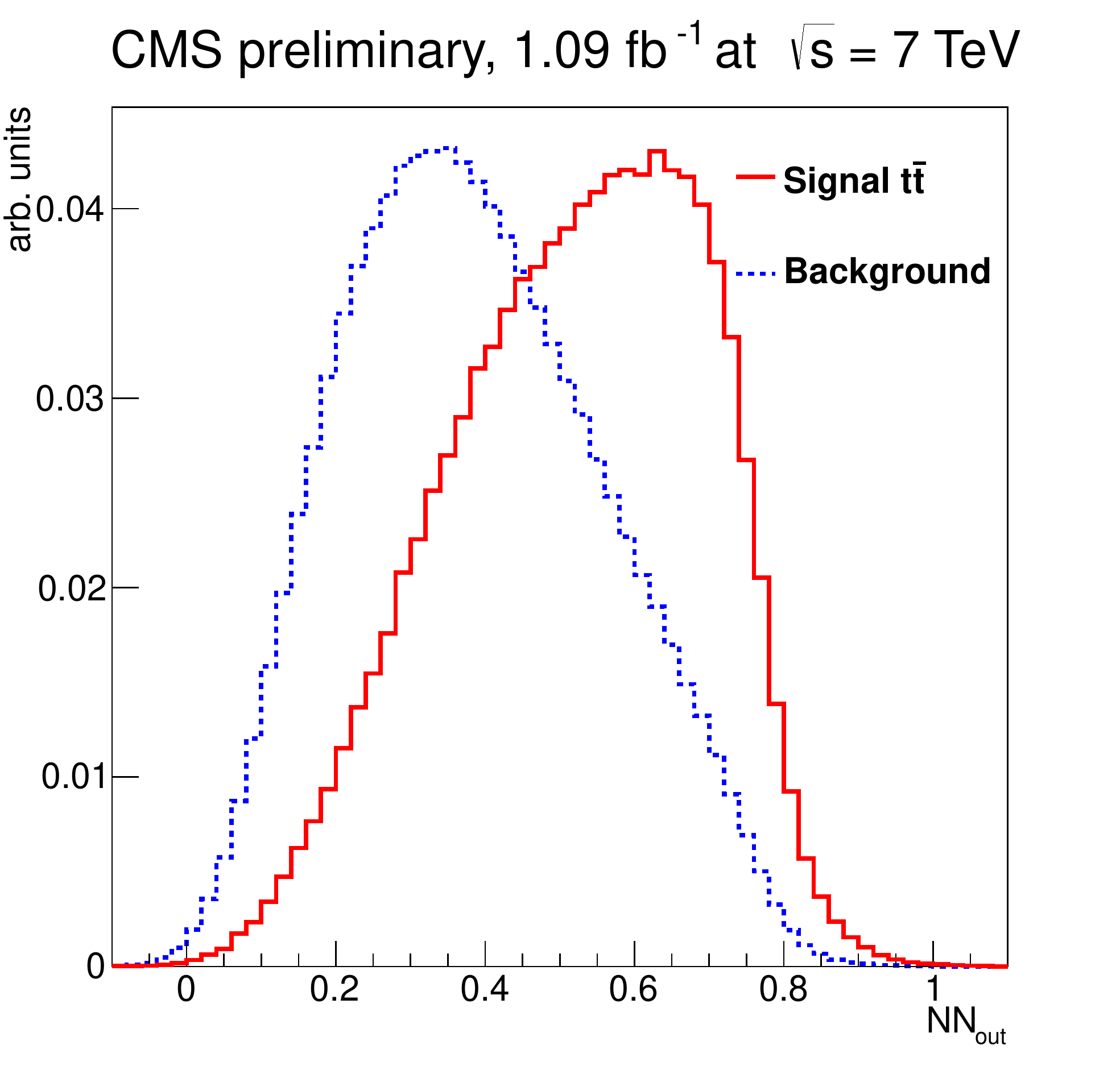}}
\caption{Neural network output for simulated $t\bar t$ events and multi-jet events from data shown normalized to unity.}
\label{fig:nn}       
\end{center}
\end{figure}

The neural network has been trained using in total eight kinematic variables. These variables are sphericity, centrality, aplanarity, the sum of transverse jet energies (including and without the first two jets), the transverse jet energies weighted with $\sin^2\theta$ ($\theta$ being the angle enclosed by the jet and the beam-axis in the center-of-mass frame of the multijet final state) for the first and second jets, as well as the mean of the remaining jets.  The events are all required to have a jet multiplicity of 
$6 \leq N_{jet} \leq 8$, where the jets selection criteria are the same as for the main analysis. The outcome of the training process is shown in Figure \ref{fig:nn}. The selected events are required to have a neural network output $NN_{out} > 0.65$.

To improve the purity of the selected sample two $b$-tags are required using an algorithm different from the on used in the main analysis, with an efficiency of 63 $\pm$ 4\% and mis-tag rate of 1.8 $\pm$ 0.2\%. The assumption for the data-driven QCD multijet background estimate is that the $b\bar b$ dynamics in events with six or more jets can be appropriately inferred from a control sample of events with four or five jets. Three variables are used to parametrize the probability that both jets in a given dijet pair are both $b$-tagged: the average transverse momentum $<p_T>$, the average pseudorapidity $<|\eta|>$, and the opening angle between the two jets $\Delta R$.
 
Events with six or more jets have by definition more dijet combinations than events with four or five jets. In order to avoid a bias due to this fact the probability that a given jet pair with both jets having a loose $b$-tag also has both jets with a medium $b$-tag was parametrized. With this parametrization, referred to as the ratio $R^{MM}_{LL}$, the expected amount of background pairs of medium-tagged jets is obtained by weighting each event with 
\begin{equation}
w = \sum R^{MM}_{LL}(<p_T>, <|\eta|>, \Delta R)                                                  
\end{equation}
where the sum runs over all loose-tagged jet pairs in the event with at least two loose $b$-tags. The $t\bar t$ signal is extracted from the reconstructed top quark mass. For the event reconstruction a kinematic fit is used, similar to the one used for the main analysis, with the requirements of $\chi^2 < 40$ and a minimum opening angle between the $b$-tagged jets of $\Delta R >$ 1.2. After the full event selection 937 events remain from which 1125 top quark masses can be reconstructed. Three templates are used to describe the reconstructed top quark mass distribution.

\begin{figure}[ht!]
\begin{center}
\resizebox{0.75\columnwidth}{!}{\includegraphics{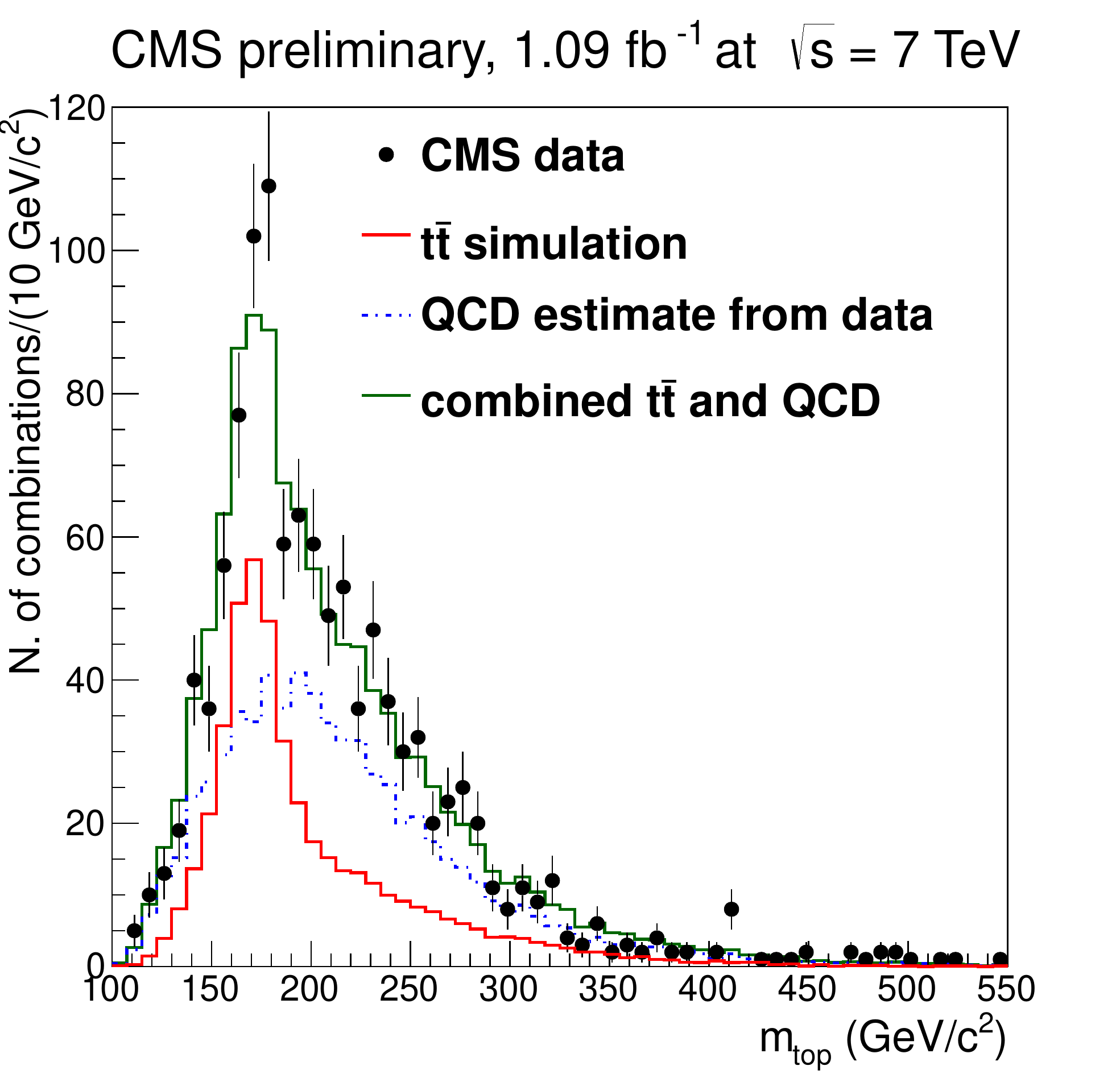}}
\caption{Reconstructed top quark mass $m_{top}$ for $b$-tagged jet combinations observed in data (circles) passing the full event selection. For comparison, the expected background (dashed line) and $t\bar t$ signal (solid line) are shown, both normalized to the yields from the fit.}
\label{fig:topmass_nn}       
\end{center}
\end{figure}

The background template is obtained from data by weighting it with the double-tag probability. The signal template is acquired directly from simulation. A third template representing signal events behaving as background is also derived from simulation, but
weighted with the double-tag probability. This is done in order to correct for contamination from signal events in the control region. Finally, the cross section is extracted from a binned maximum likelihood fit of these three templates to the distribution measured from data. Figure \ref{fig:topmass_nn} shows the comparison between the reconstructed top mass from data and the expected signal and background distributions normalized to the yields found in the fit. The measured cross section is  $\sigma_{t \bar t}= 157 \pm 30 (stat.) \pm 47 (sys.) \pm 8 (lumi.)$ pb with a signal fraction of 40\%.

The value is in agreement with the Standard Model predictions and with the value obtained by the main analysis. This provides an important and independent cross check of the measurement. 

\section{Conclusions}
A first measurement of the top quark pair production cross section in the fully hadronic decay channel at a center-of-mass energy of 7 TeV has been presented. The measurement results in a cross section of
\begin{equation}
\sigma_{t \bar t}= 136 \pm 20 (stat.) \pm 40 (sys.) \pm 8 (lumi.) 	\rm{~ pb}
\end{equation}

A cross check analysis, using a neural network based event selection and a different QCD multijet background estimate, gives a result compatible within the errors. Both results are consistent with earlier CMS measurements in the dilepton and lepton+jets decay channels 
and also with the Standard Model approximate NNLO theoretical calculations \cite{kidonakis,ahrens,athor}.

\end{document}